\documentclass[a4paper,11pt]{article}%
\usepackage{amsmath,amssymb,mathrsfs}
\usepackage{latexsym,graphicx,graphics,multirow,fancyhdr,time,color,bm,tabularx,psfrag}
\usepackage{geometry}
\begin{document}
\title{Holographic multi-condensate with nonlinear terms}
\author{Xing-Kun Zhang,~Chuan-Yin Xia,~
Zhang-Yu Nie\thanks{niezy@kust.edu.cn,~Corresponding author.},~
Hui Zeng\thanks{zenghui@kust.edu.cn,~Corresponding author.}}
\maketitle
\abstract{We study the influence of nonlinear terms quartic of the charged fields, which do not change the critical points of single condensate solutions, on the phase structure of a holographic model with multi-condensate in probe limit. We include one s-wave order and one p-wave order charged under the same U(1) gauge field in the holographic model and study the influence of the three quartic nonlinear terms of the charged fields with coefficients $\lambda_s$, $\lambda_p$ and $\lambda_{sp}$ on the phase structure. We show the influence of each of the three parameters on the phase diagram with other two set to zero, respectively. With these nonlinear terms, we get more power on tuning the phase structure of the holographic system showing multi-condensate, and show how to get a reentrant phase transition as an example.
}
\section{Introduction}
The holographic superconductor models with single condensate~\cite{Hartnoll:2008vx,Gubser:2008wv,Chen:2010mk,Benini:2010pr,Cai:2013aca} has been extended to systems with multiple orders~\cite{Cai:2015cya} in recent years. The competition and coexistence of two orders are firstly studied in a holographic s+s model~\cite{Basu:2010fa,Cai:2013wma}, and later in s+p~\cite{Nie:2013sda,Nie:2014qma,Amado:2013lia,Arias:2016nww} and s+d~\cite{Nishida:2014lta,Li:2014wca} models. In these models, besides the single condensate solutions, new coexistent solutions also exist, as a result, the holographic models with multi-condensate show rich phase transitions and the phase structures become more complicated.

The holographic models with multi-condensate show rich phase transitions even in probe limit, which is quite useful for further studies involving time evolution. For example, in Ref.~\cite{Yang:2019ibe}, phase separation are realized in a rotating system with two s-wave condensates holographically in probe limit. In Ref.~\cite{Li:2020ayr}, the authors use a similar model with two s-wave orders to get a first order phase transition in probe limit, and show the formation of two dimensional bubble structure in first order phase transitions holographically. In the holographic study of time dependent physics~\cite{Bhaseen:2012gg,Adams:2012pj,Adams:2013vsa,Sonner:2014tca,Chesler:2014gya}, taking the probe limit help avoid the complicated numerical relativity problems. Therefore, exploring the phase structure of the holographic models with multiple condensates in probe limit are important and benefits future studies.

In the study of the phase structure of the holographic models, the influence of various parameters on the critical points are quite crucial. The quartic nonlinear terms such as $\lambda |\Psi|^4$ may change the phase transition from a second order one to a first order one~\cite{Herzog:2010vz}, but do not change position of the critical point which is determined by the physics in linear region. However, in systems with multiple condensate, the critical points of the coexistent solution are emerging from the single condensate solution with finite condensate value, as a result, the phase structure involving the coexistent solution get non-trivial effects from such nonlinear terms. Therefore, it is interesting to study the influence of the quartic nonlinear terms on the phase structure of multi-condensate holographic models.

It is difficult to build holographic models dual to a special material, therefore the recent study mainly focus on possible universality in the strongly coupled systems, where a first step is usually realizing specific phase transitions and various phase diagrams~\cite{Nie:2015zia,Kiritsis:2015hoa,Chen:2016cym}. Thus it is useful to investigate skills and technics to build various kind of phase transitions in holography. One kind of interesting phase transition is the reentrant one, which exhibit non-monotonic dependent of the order parameter, and has already been realized holographically. However, the reentrant phase transition is usually studied in a holographic model with considering back reaction on metric~\cite{Nie:2014qma}, or involving physical problems such as causality violation~\cite{Li:2017wbi}, which block the further investigations. With the nonlinear parameters turned on, we get more power on tuning the phase transitions. These nonlinear terms have different influence as that of the other parameters such as charge $q$ and mass $m^2$, which change the thermodynamic potential curve ``parallel''~\cite{Li:2017wbi}.



In this paper, we study the influence of three quartic nonlinear terms on the phase structure of a holographic s+p model in probe limit, and build a reentrant phase transition to test the power of tuning the phase transitions. In Section.~\ref{sect:setup} we give the basic setup of the holographic model, and show details of calculation. In Section.~\ref{sect:tuning}, we show how the interactions terms influence the phase structure of the holographic system with multi-condensate, and finally show how to use the new power to get a reentrant phase transition as a example. We conclude and discuss on our results and future work in Section.~\ref{sect:conclusion}.


\section{Holographic model of an S+P superconduction}  \label{sect:setup}
\subsection{The model setup}
We consider a holographic model with one s-wave order and one p-wave order from (3+1) dimensional asymptotic AdS spacetime. The action of this holographic s+p model is
\begin{eqnarray}
S&=&S_{M}+S_{G}~,\\
S_G&=&\frac{1}{2\kappa_g ^2}\int d^{4}x\sqrt{-g}(R-2\Lambda)~,
\\
S_M&=&\int d^{4}x\sqrt{-g}\Big(-\frac{1}{4}F_{\mu\nu}F^{\mu\nu}
-D_{\mu}\psi^{\ast}D^{\mu}\psi-m^{2}_{s}\psi^{\ast}\psi \nonumber \\ \nonumber &&  \quad\quad\quad
-\frac{1}{2}\rho^{\dagger}_{\mu\nu}\rho^
{\mu\nu}-m^{2}_{p}\rho^{\dagger}_{\mu}\rho^{\mu}\\  &&  \quad\quad\quad
-\lambda_{s}(\psi^{\ast}\psi)^{2}-\lambda_{p}(\rho^{\dagger}_{\mu}\rho^{\mu})^{2}
-\lambda_{sp}\psi^{\ast}\psi\rho^{\dagger}_{\mu}\rho^{\mu}\Big)~,
\end{eqnarray}
where $\rho_{\mu\nu}=\bar{D}_{\mu}\rho_{\mu}-\bar{D}_{\nu}\rho_{\mu}$ with the covariant derivatives $\bar{D}_{\mu}=\nabla_{\mu}-iq_{p}A_{\mu}$ and
$D_{\mu}\psi=\nabla_{\mu}\psi-iq_{s}A_{\mu}\psi$. $F_{\mu\nu}=\nabla_{\mu}A_{\nu}-\nabla_{\nu} A_{\mu}$ is the Maxwell field strength. $\Lambda=-3/L^2$ is the negative cosmological constant and $L$ is the AdS radius.

In this study, we work in the probe limit and choose the background geometry to be a 3+1 dimensional black brane with the metric
\begin{eqnarray}
ds^{2}=-f(r)dt^{2}+\frac{1}{f(r)}dr^{2}+r^{2}dx^{2}+r^{2}dy^{2},
\end{eqnarray}
where the function $f(r)$ is given by
\begin{eqnarray}
f(r)=r^{2}(1-(\frac{r_{h}}{r})^3)~,
\end{eqnarray}
with $r_{h}$ the horizon radius. The Hawking temperature of this black brane solution is given by
\begin{eqnarray}
T= \frac{3 r_h}{4\pi L^2}.
\end{eqnarray}

The consistent ansatz for the matter fields can be taken as
\begin{eqnarray}
\psi=\psi_{s}(r)~, A_{t}=\phi(r)~, A_{x}=\psi_{p}(r)~,
\end{eqnarray}
and all other field components are set to zero. Consequently, the equations of motion of matter fields are
\begin{eqnarray}
\phi''+\frac{2}{r}\phi'-2(\frac{q^{2}_{s}\psi^{2}_{s}}{f}+\frac{q^{2}_{p}L^{2}\psi^{2}_{p}}{r^{2}f})\phi=0~,\label{EqPhi}\\
\psi_{p}''+\frac{f'}{f}\psi_{p}'+(\frac{q_{p}^{2}\phi^{2}}{f^{2}}-\frac{m^{2}_{p}}{f}-\frac{\lambda_{sp}\psi_{s}^{2}}{f})\psi_{p}-\frac{2\lambda_{p}L^{2}}{r^{2}f}\psi^{3}_{p}=0~,
 \label{EqPsip}\\
\psi_{s}''+(\frac{f'}{f}+\frac{2}{r})\psi_{s}'+(\frac{q_{s}^{2}\phi^{2}}{f^{2}}-\frac{m^{2}_{s}}{f}-
\frac{\lambda_{sp}L^{2}\psi_{p}^{2}}{r^{2}f})\psi_{s}-\frac{2\lambda_{s}}{f}\psi^{3}_{s}=0~.\label{EqPsis}
\end{eqnarray}
The three terms in the action $-\lambda_{s}(\psi^{\ast}\psi)^{2}$, $-\lambda_{p}(\rho^{\dagger}_{\mu}\rho^{\mu})^{2}$, $-\lambda_{sp}\psi^{\ast}\psi\rho^{\dagger}_{\mu}\rho^{\mu}$ are quartic to the charged fields $\psi_s$ and $\psi_p$ and introduce nonlinear terms to the two equations (\ref{EqPsip}), (\ref{EqPsis}). The nonlinear property of the terms in single equation is important to understand the influence on critical points. Therefore we call these terms nonlinear in the sense that the single equation (\ref{EqPsip}) or (\ref{EqPsis}) is not linear on the condensate fields $\psi_p$ and $\psi_s$, although the equations of motion are already not linear as a whole even with out these quartic terms.
\subsection{Boundary conditions}
To solve the equations of motion, we need boundary conditions both on the horizon and on the boundary.
The boundary expansions of the three fields on the horizon side are
\begin{eqnarray}
&\phi(r)=\phi_{1}(r-r_{h})+\mathcal{O}((r-r_{h})^{2})~,\\
&\psi_{s}(r)=\psi_{s0}+\mathcal{O}(r-r_{h})~,\\
&\psi_{p}(r)=\psi_{p0}+\mathcal{O}(r-r_{h})~,
\end{eqnarray}
where $\phi(r=r_h)$ is set to zero to meet the physical constraints and the equations of motion (\ref{EqPsip}), (\ref{EqPsis}) imply additional constraints on the solutions with finite value at the horizon. Therefore, only $\phi_{1}$,$\psi_{s0}$,$\psi_{p0}$ are independent parameters.
The expansions near the AdS boundary are
\begin{eqnarray}
&\phi(r)=\mu-\frac{\rho}{r^2}+...~,\\
&\psi_{p}=\psi_{p_{-}}+\frac{\psi_{p_{+}}}{r^2}+...~,\\
&\psi_{s}=\frac{\psi_{s_{-}}}{r^{\Delta_{-}}}+\frac{\psi_{s_{+}}}{r^{\Delta_{+}}}+...~,
\end{eqnarray}
where
\begin{equation}
\Delta_{s\pm}=\frac{(3\pm\sqrt{9+4m_{s}^{2}})}{2}~,
\Delta_{p\pm}=\frac{(1\pm\sqrt{1+4m_{p}^{2}})}{2}~.
\end{equation}
$\mu$ and $\rho$ are the chemical potential and charge density, respectively.
We choose standard quantization, which means that $\psi_{s-}$ and $\psi_{p-}$ are the sources of the s-wave and p-wave orders, while $\psi_{s+}$ and $\psi_{p+}$ are the expectation values.
The three degrees of freedom are fixed with $\psi_{s-}=\psi_{p-}=0$ and the value of $\mu$.

\subsection{Grand potential}
In order to compare the stability of the different solutions and get the final phase diagram, we study in the grand canonical ensemble and calculate the grand potential to find the stability relation between different solutions.
The grand potential of the system can be identified with the temperature times the Euclidean on-shell action of the bulk solution according to the AdS/CFT correspondence.
In this paper we work in probe limit, therefore, the difference of the grand potential only come from the matter part of the action
\begin{eqnarray}
\Omega_{m}=T S_{ME},
\end{eqnarray}
where $T$ is the Hawking temperature and $S_{ME}$ is the value of the matter action calculated with Euclidean time. $\Omega_m$ is the contribution from the matter action to the grand potential and with the equations of motion been substituted in, can be expresses as
\begin{eqnarray}
\Omega_{m}=&\displaystyle{\frac{V_{2}}{T}}\Big(-\displaystyle{\frac{\mu\rho}{2L^2}}
+&\int^{\infty}_{r_h}\big(\frac{q_{s}^{2}r^{2}\phi^{2}\psi_{s}^{2}}{L^{2}f}
+\frac{q_{p}^{2}\phi^{2}\psi_{p}^{2}}{f}  \nonumber \\  &&
-\frac{\lambda_{s}r^{2}\psi_{s}^{4}}{L^{2}}-\frac{\lambda_{p}L^{2}\psi_{p}^{4}}{r^{2}}
-\lambda_{sp}\psi_{s}^{2}\psi_{p}^{2}\big)dr\Big).
\end{eqnarray}

\subsection{scaling symmetry}
There are three sets of scaling symmetries in the Equations (\ref{EqPhi}),(\ref{EqPsip}) and (\ref{EqPsis})
\begin{eqnarray}
&\phi\rightarrow \lambda\phi,\psi_s\rightarrow\lambda\psi_s,\psi_p\rightarrow\lambda\psi_p,
q_s\rightarrow\lambda^{-1} q_s,q_p\rightarrow\lambda^{-1} q_p,\nonumber\\
&\lambda_s\rightarrow\lambda^{-2}\lambda_s, \lambda_p\rightarrow\lambda^{-2}\lambda_p,\lambda_{sp}\rightarrow\lambda^{-2}\lambda_{sp}; \label{scaling qs}
\\
&\phi\rightarrow \lambda^{-2}\phi,\psi_s\rightarrow\lambda^{-1}\psi_s,f\rightarrow\lambda^{-2}f,
L\rightarrow\lambda^{-1}L,\nonumber\\
&m_s^2\rightarrow\lambda^{-2}m_s^2, \label{scaling L},m_p^2\rightarrow\lambda^{-2}m_p^2;
\\
&\phi\rightarrow\lambda^{-1}\phi,\psi_p\rightarrow\lambda^{-1}\psi_p,
f\rightarrow\lambda^{-2}f,r\rightarrow\lambda^{-1}r,r_h\rightarrow\lambda^{-1}r_h.
\label{scaling r}
\end{eqnarray}
With the second and third scaling symmetries, we set $r_h=L=1$ for numerical computation. After we get the numerical solutions, we can use again the two sets of scaling symmetries to recover $r_h$ and $L$ to any value.

\section{Tuning the parameters $\lambda_{s}$, $\lambda_{p}$, $\lambda_{sp}$ and phase diagram}\label{sect:tuning} 
There are three different solutions with non-zero condensates, called the s-wave solution, the p-wave solution and the s+p solution, respectively. It is easy to get the s-wave and p-wave solutions with single condensate. However, we need more technic to get the s+p solution. It is conjectured that the s+p solution always exist in a narrow region near the intersection point of the grand potential curves of s-wave and p-wave solutions in probe limit~\cite{Nie:2014qma}. Therefore, we need to first locate this intersection point to get the s+p solution.

To focus on the influence of the three quartic terms with parameters $\lambda_s$, $\lambda_p$ and $\lambda_{sp}$, we firstly fix the value of the other parameters ${m_s^2,~m_p^2}$ and ${q_s,~q_p}$. In the study of time dependent evolution in holographic superconductors, it is wise to set the dimension $\Delta_{s}, \Delta_p$ to integer value. And a best choice for the mass parameters with $m_s^2=-2$ and $m_p^2=0$ further simplify the equations. 

Because of the first scaling symmetry~(\ref{scaling qs}), only the ratio $q_p/q_s$ changes the qualitative feature of the phase diagram. Therefore we set $q_s=1$ without lose of generality. Furthermore, we tune $q_p$ to make the grand potential of the s-wave solution and the p-wave solution have one intersection point when $\lambda_{s}=\lambda_{p}=\lambda_{sp}=0$, which indicate the existence of the coexistent solution. We should notice that when we tune $q_p$, the grand potential curve for the p-wave solution move ``parallel''~\cite{Li:2017wbi}.

At last, we fix ${m_s^2=-2,~m_p^2=0}$ and ${q_s=1,~q_p=0.8881}$ to focus on the influence of the three parameters $\lambda_s$, $\lambda_p$ and $\lambda_{sp}$ for the non linear terms. We start from the case with $\lambda_{s}=\lambda_{p}=\lambda_{sp}=0$ as a reference.

\subsection{Grand potential and condensates for $\lambda_{s}=\lambda_{p}=\lambda_{sp}=0$}
\begin{figure}
\center
\includegraphics[width=0.47\columnwidth,origin=c,trim=60 260 120 270,clip]{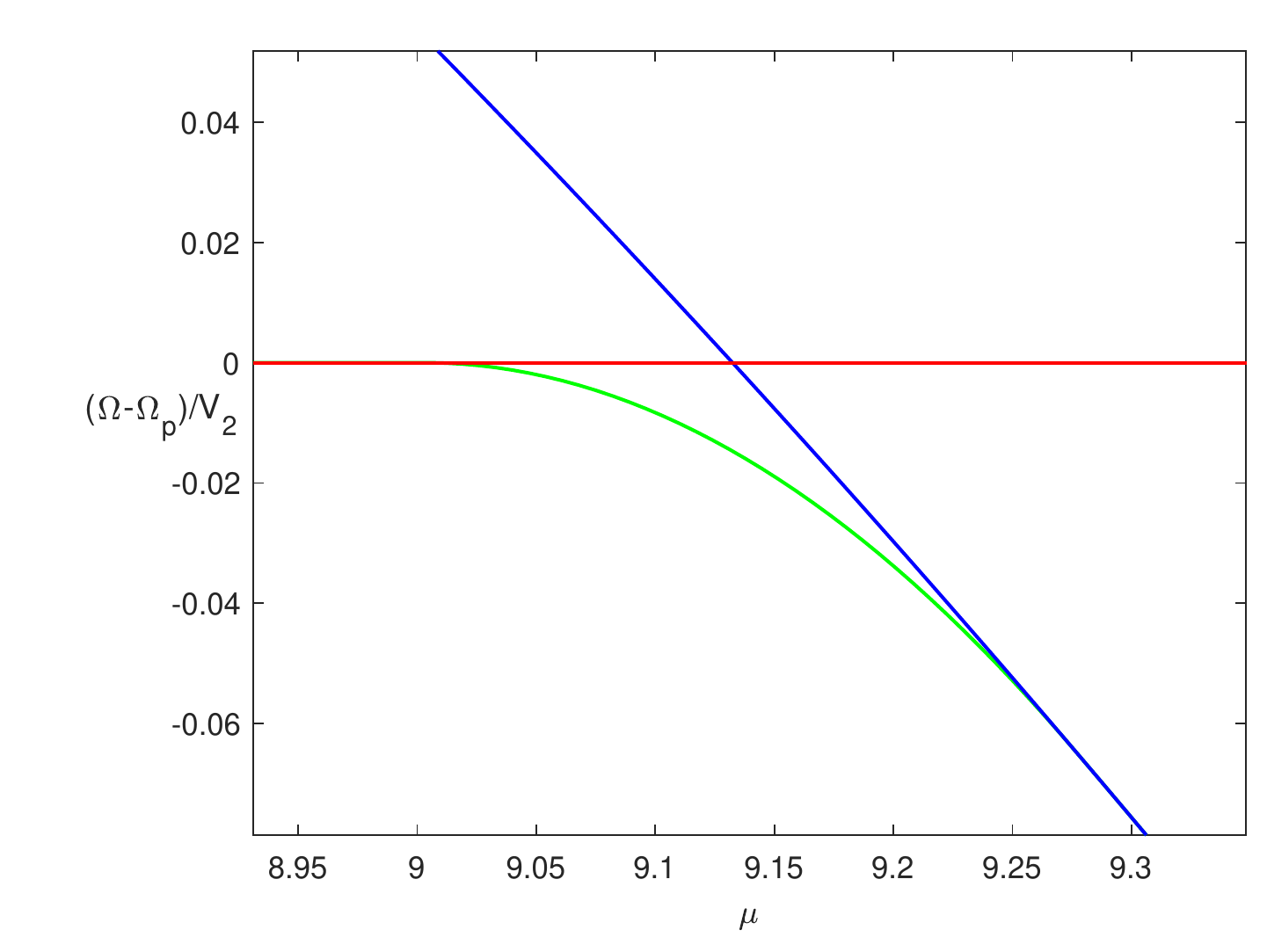}
\includegraphics[width=0.47\columnwidth,origin=c,trim=60 260 120 270,clip]{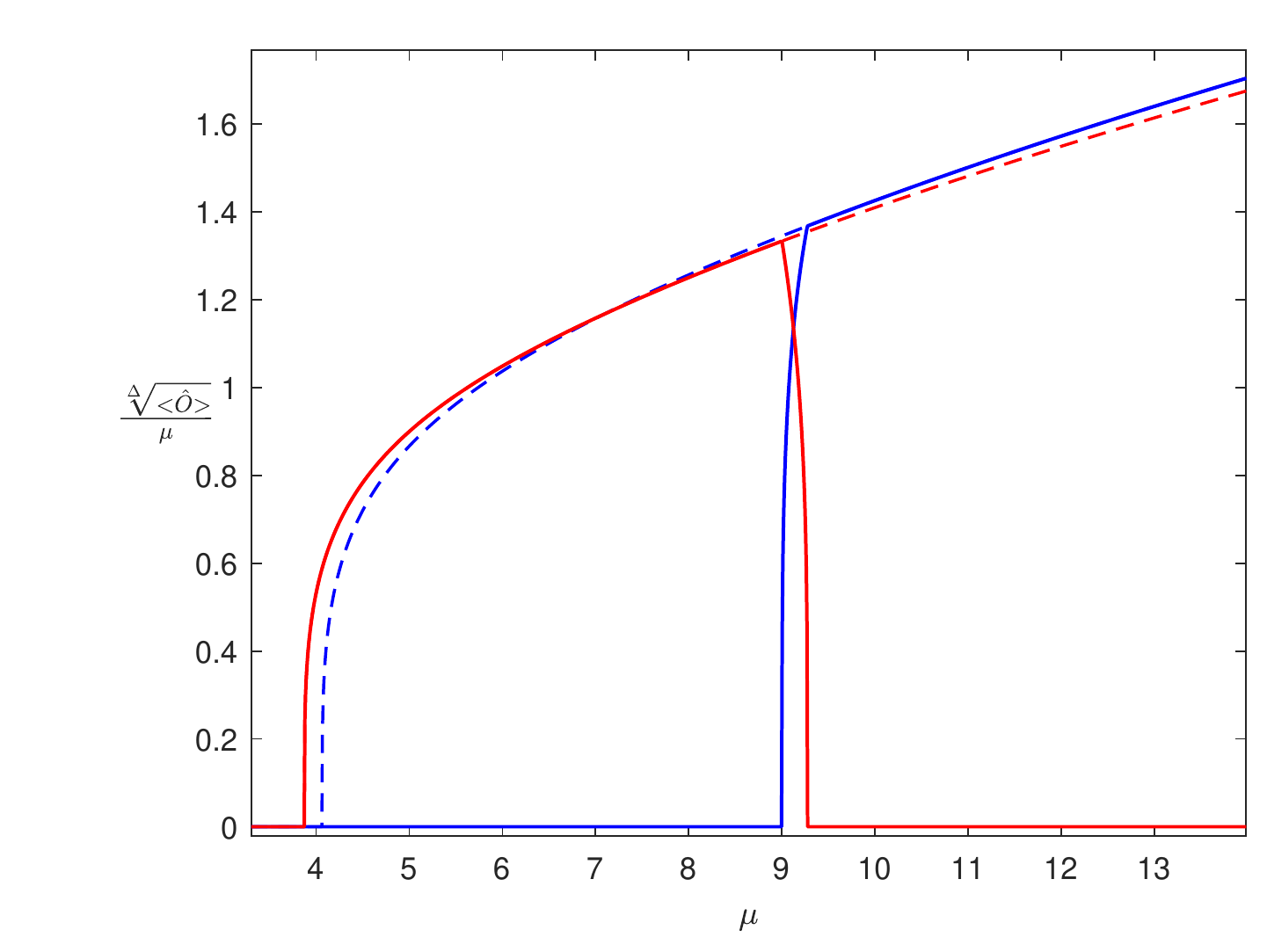}
\caption{Grand potential(left plot) and condensates(right plot) for different solutions. In the left plot, the three curves show the relative value of grand potential density with respect to the p-wave solution $(\Omega-\Omega_p)/V_2$ for the s-wave solution(blue line), the p-wave solution(red line) and the s+p solution(green line), respectively. In the right plot, the blue and red lines denotes the condensate value for the s-wave and p-wave orders respectively. The solid lines show the condensate values for the most stable solutions while the dashed lines show the condensate values for the unstable region of the s-wave and p-wave solutions.
}\label{Fig1}
\end{figure}

We start from the case with ($\lambda_{s}=\lambda_{p}=\lambda{sp}=0$) as a reference, where the grand potential curves of the s-wave solution and the p-wave solution have an intersection point. Near this intersection point, we can find the s+p solution, which has the lowest value of grand potential. We show the relative value of grand potential as well as the condensates in Figure.~\ref{Fig1}.

In the left plot of Figure.~\ref{Fig1}, we show the value of grand potential for different solutions with respect to the grand potential of the p-wave solution. We can see that the s+p solution emerge from the p-wave solution, and then leave away and merge with s-wave solution finally. In the right plot, we show the condensate value of the two orders for different solutions. 
We can see that at the left critical point, the s+p solution has an infinitesimal s-wave condensate, while at the right critical point, the s+p solution has an infinitesimal p-wave condensate. Therefore, in order to make the reference to the two critical points more clearly, we call the critical point with an infinitesimal s-wave condensate as the critical point for the s-wave order of the s+p solution, and call the critical point with an infinitesimal p-wave condensate as the critical point for the p-wave order of the s+p solution. Both the two critical points in Figure~\ref{Fig1} are second order, but the phase transitions may become first order when the three quartic terms are turned on. For convenience, when the phase transition become first order, we still use the word ``critical point'' to denote the spinodal point in first order phase transition, which also get an infinitesimal value of one condensate.

\subsection{Tuning $\lambda_s$, $\lambda_p$}
Base on the case with $\lambda_{s}=\lambda_{p}=\lambda_{sp}=0$ shown in the previous subsection, we further change $\lambda_s$ and $\lambda_p$ respectively to tune the phase structure of the system, and show the resulting $\lambda_s-\mu$ and $\lambda_p-\mu$ phase diagrams, respectively.

We firstly study the influence of the two parameters on the single condensate solutions. From the equations of motion, we can see that $\lambda_s$ does not change the p-wave solution with $\psi_s=0$, and $\lambda_p$ does not change the s-wave solution with $\psi_p=0$ either. Therefore we can focus on the influence of $\lambda_s$ on the s-wave solution and the influence of $\lambda_p$ on the p-wave solution.

\begin{figure}
  \centering
\includegraphics[width=0.49\columnwidth,origin=c,trim=90 260 120 230,clip]{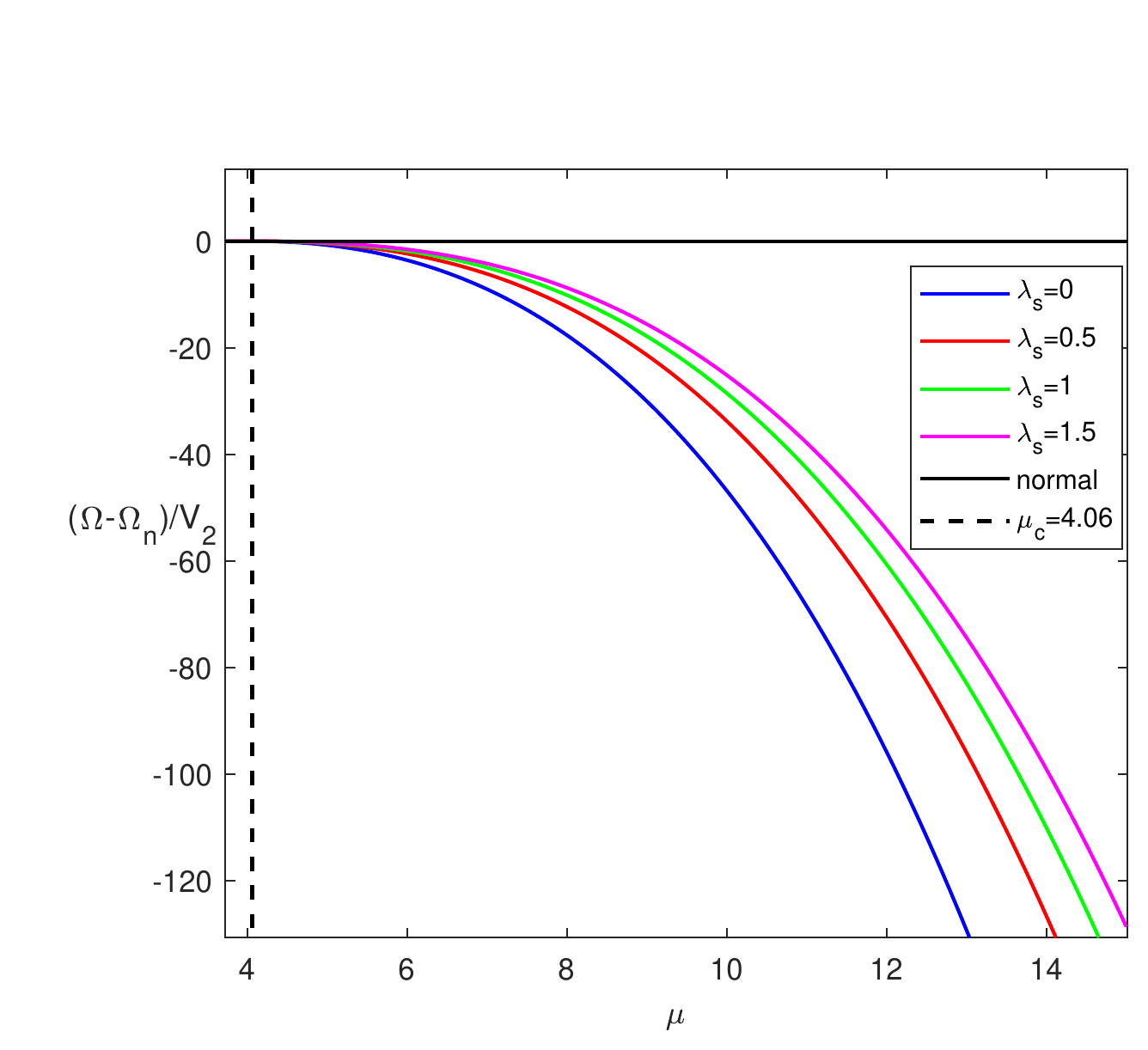}
\includegraphics[width=0.49\columnwidth,origin=c,trim=90 260 120 230,clip]{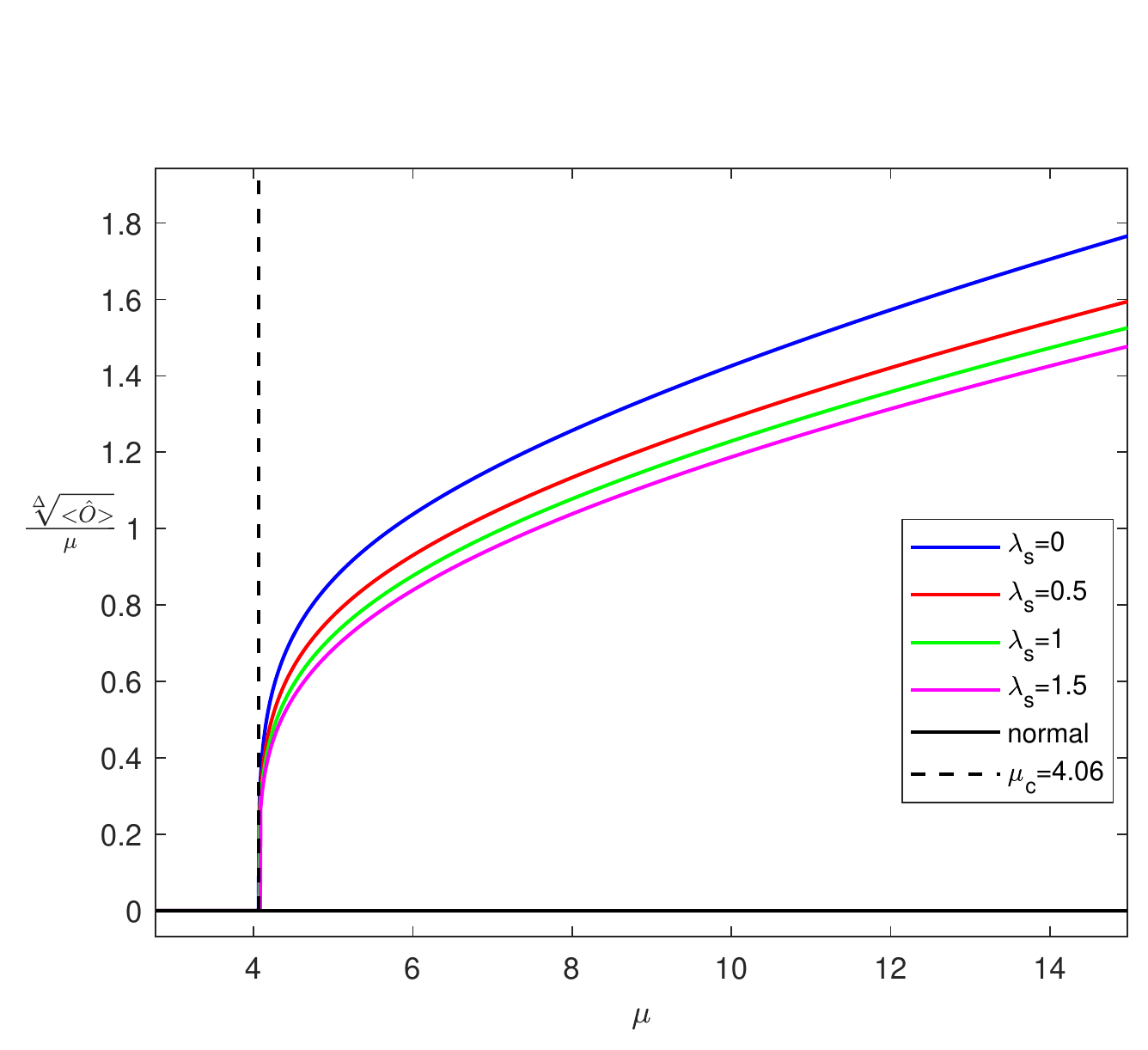}
\caption{The relative value of grand potential density for the s-wave solutions with respect to the normal solution(left plot) and the condensates of s-wave order for the s-wave solutions(right plot) with $\lambda_s=0$(blue), $\lambda_s=0.5$(red), $\lambda_s=1$(green) and $\lambda_s=1.5($purplish red), respectively. The dashed black line represents the position of the critical point which is not influenced by the value of $\lambda_s$.}\label{fig2}
\end{figure}

We show the grand potential as well as condensates of the s-wave solutions with different values of $\lambda_s$ in Figure.~\ref{fig2}. We can see that the value of $\lambda_s$ does not change the critical point. This is because that the term with $\lambda_s$ is nonlinear for $\psi_s$ in the single equation (\ref{EqPsis}), and does not change the critical behavior determined by infinitesimal value of $\psi_s$. Further more, it is obvious that the term with coefficient $\lambda_s$ have stronger influence on the s-wave solution with larger value of condensate. This could also be explained by the cubic dependence on $\psi_s$ of the term $\lambda_s \psi_s^3$ in the equations of motion.

\begin{figure}
  \centering
\includegraphics[width=0.49\columnwidth,origin=c,trim=90 260 120 230,clip]{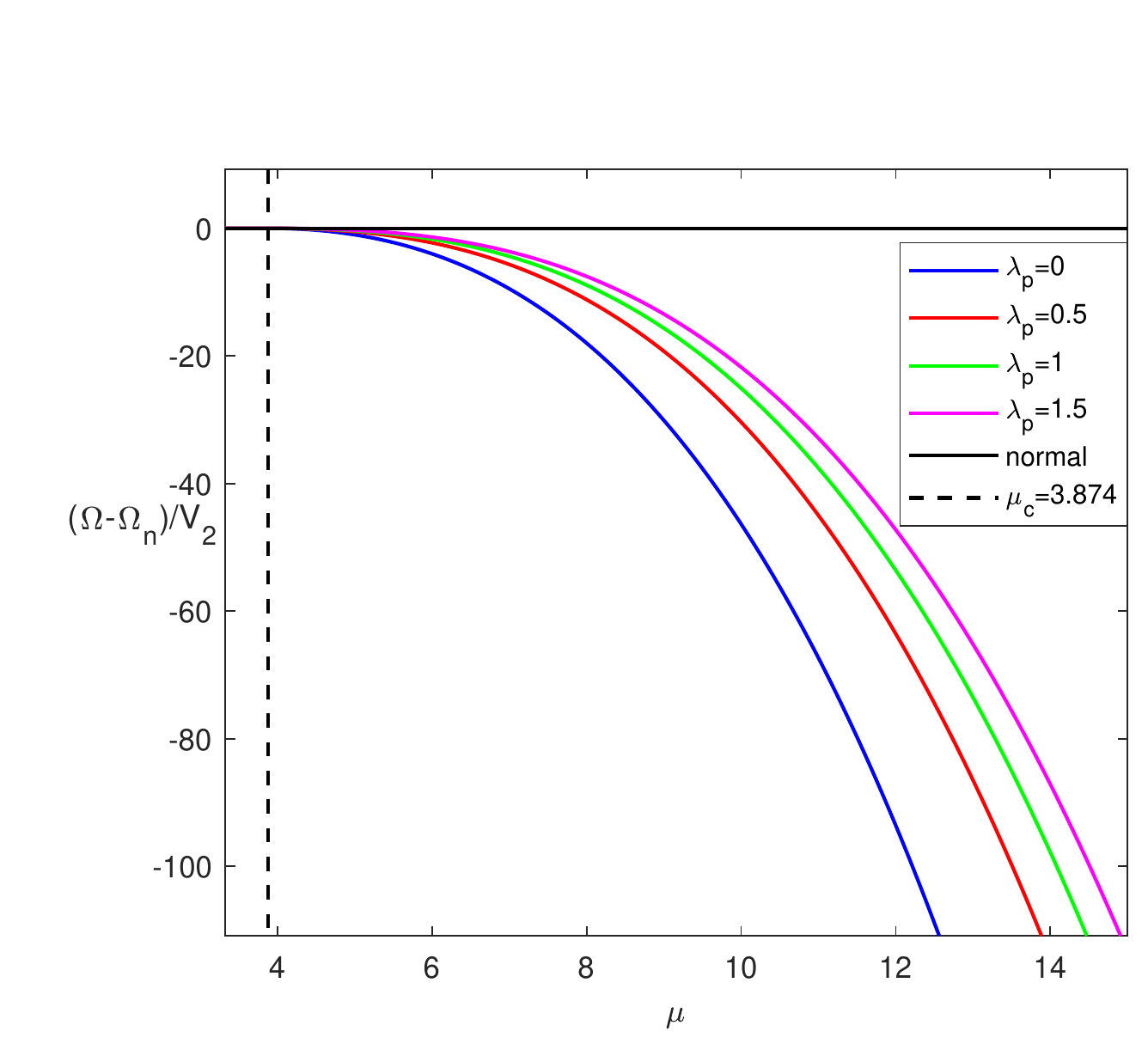}
\includegraphics[width=0.49\columnwidth,origin=c,trim=90 260 120 230,clip]{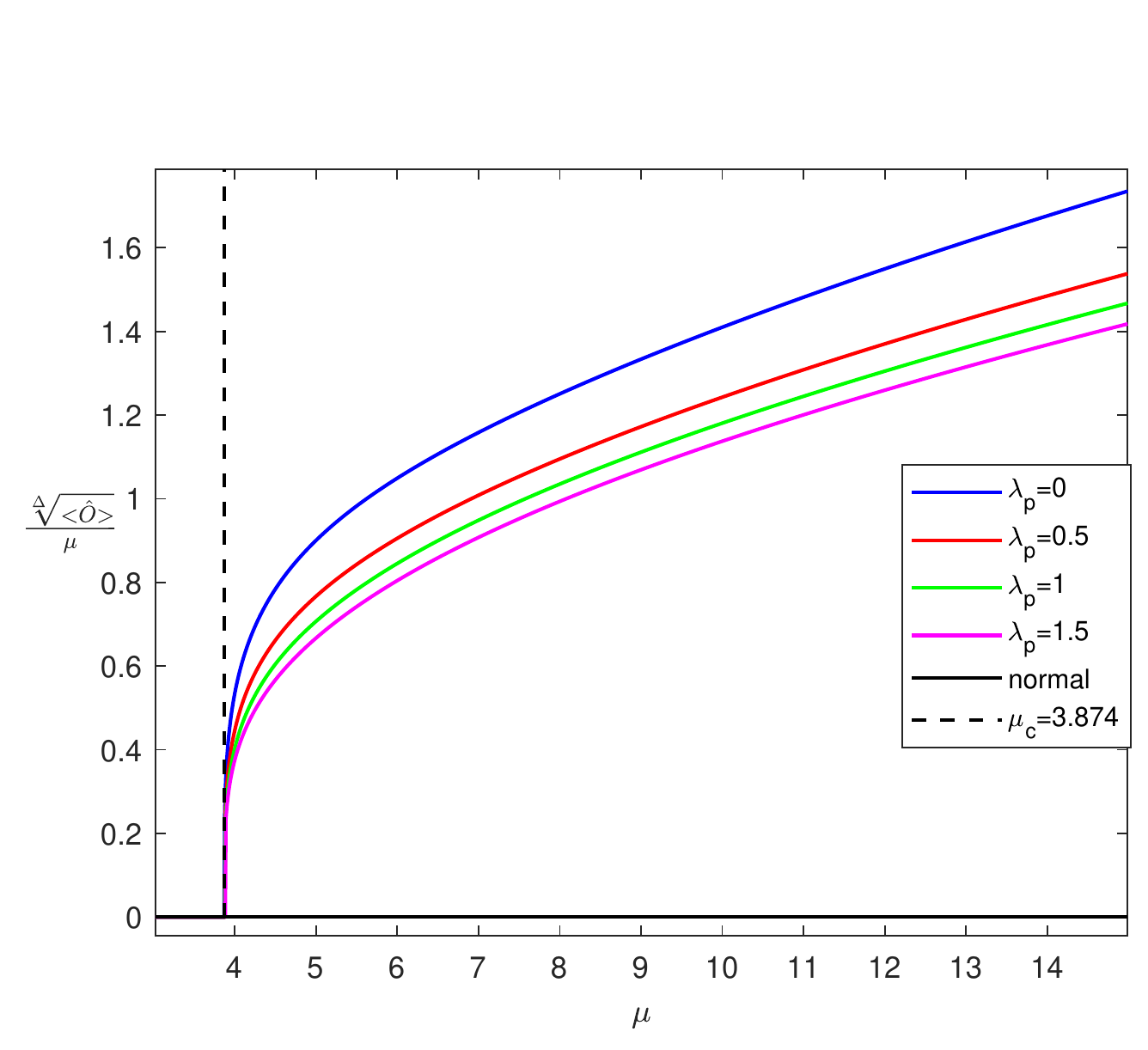}
\caption{The relative value of grand potential density for the p-wave solutions with respect to the normal solution(left plot) and the condensates of p-wave order for the p-wave solutions(right plot) with $\lambda_p=0$(blue), $\lambda_p=0.5$(red), $\lambda_p=1$(green) and $\lambda_s=1.5($purplish red), respectively. The dashed black line represents the position of the critical point which is not influenced by the value of $\lambda_p$.}\label{fig3}
\end{figure}

The influence of the value of $\lambda_p$ on the p-wave solution is qualitatively the same to the influence of the value of $\lambda_s$ on the s-wave solution. We show the similar results of the grand potential as well as condensates for the p-wave solution with different values of $\lambda_p$ in Figure.~\ref{fig3}.

Next, we study the influence of the two parameters on the coexistent s+p solution. Again there are similarities between the influence of the values of $\lambda_s$ and $\lambda_p$, therefore we can focus on analysis of the case for $\lambda_s$, and the case for $\lambda_p$ can be understood similarly.

\begin{figure}
  \center
\includegraphics[width=0.49\columnwidth,origin=c,trim=90 260 120 230,clip]{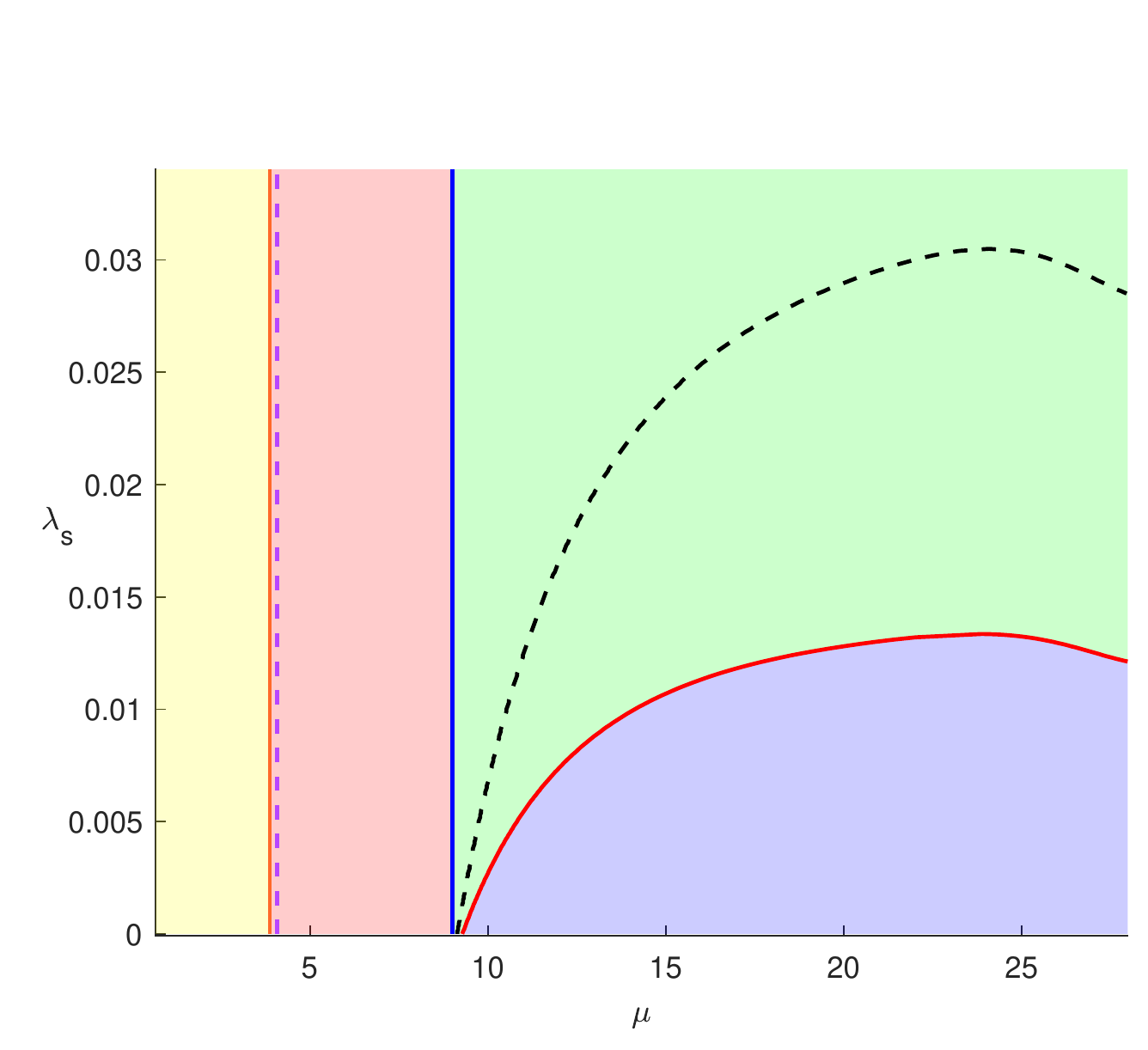}
\includegraphics[width=0.49\columnwidth,origin=c,trim=90 260 120 230,clip]{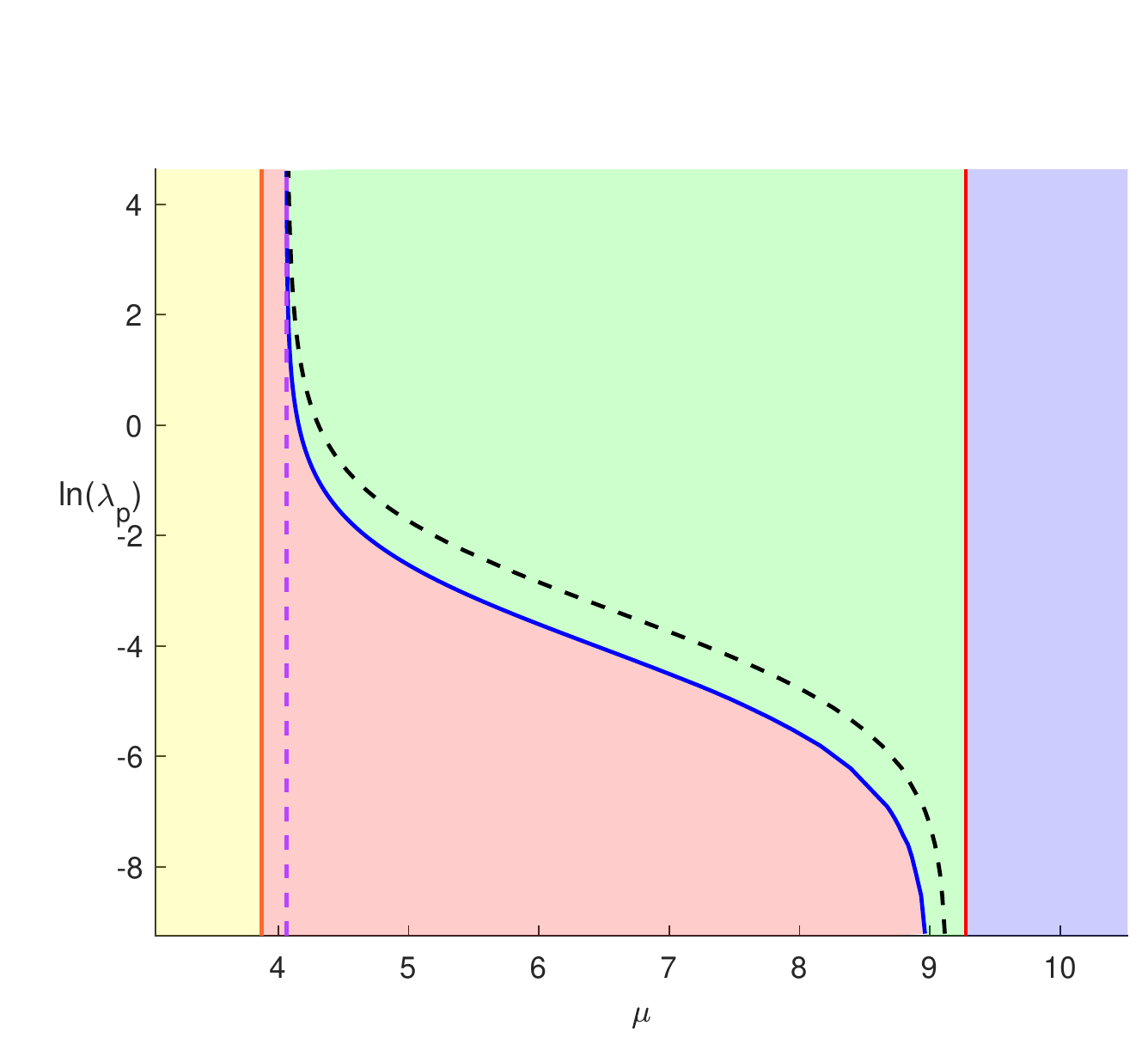}
\caption{The $\lambda_s-\mu$ phase diagram with $\lambda_p=\lambda_{sp}=0$(left plot) and the $\lambda_p-\mu$ phase diagram with $\lambda_s=\lambda_{sp}=0$(right plot). In the two phase diagrams, we use yellow, red, blue and green to denote the regions dominated by the normal phase(N), the p-wave phase(P), the s-wave phase(S) and the s+p phase(S+P), respectively. The solid orange line between the yellow and the red region presents the critical points of the p-wave solution and the dashed purple line denotes the critical points of the s-wave solution. The blue and red solid lines represent the critical points for the s-wave order and p-wave order of the s+p solution, respectively. The dashed black line represents the intersection points of grand potential curves for the s-wave solution and the p-wave solution.}\label{fig4}
\end{figure}

We show the $\lambda_s-\mu$ phase diagram with $\lambda_p=\lambda_{sp}=0$ in the left plot of Figure.~\ref{fig4}. In this plot, we use yellow, red, blue and green to denote the regions dominated by the normal phase, the p-wave phase, the s-wave phase and the s+p phase, respectively. The solid orange line between the yellow and the red region presents the critical points of the p-wave solution and the dashed purple line denotes the critical points of the s-wave solution. Because the critical chemical potential for the p-wave solution $\mu_{c-p}$ is smaller than the critical chemical potential $\mu_{c-s}$, the solid orange line is also a set of phase transition points while the dashed purple line is not. The blue and red solid lines represent the critical points for the s-wave order and p-wave order of the s+p solution, respectively. The dashed black line represents the intersection points of grand potential curves for the s-wave solution and the p-wave solution.

As we have just analyzed, different value of $\lambda_s$ do not affect the critical points of the s-wave solution and p-wave solution, therefore both the solid orange and dashed purple curves are vertical. When the value of $\lambda_s$ becomes larger, the region of the s+p phase increases as a result of the right critical point moving rightwards along the red solid line. However, the left critical point for the s-wave order do not move with the increasing of $\lambda_s$, and the blue solid line is also vertical. This property can also be deduced from analytical discussion of the equations of motion. We have pointed out that the value of $\lambda_s$ do not change the p-wave solution. Furthermore, at the critical point for s-wave order of the s+p solution, the s-wave condensate is infinitesimal. Therefore this critical point is determined by the linearized version of Equation (\ref{EqPsis}) for $\psi_s$ with function $\phi$ determined by the p-wave solution. Both the two factors are independent of $\lambda_s$, as a result, this critical point for s-wave order of the s+p solution is not dependent on $\lambda_s$, and the solid blue line is vertical in the phase diagram.

Another property of the coexistent solution is that the red solid line denoting the critical point for p-wave order of the s+p solution is not monotonic and show a maximum value of $\lambda_s=\lambda_s^m$. This implies that when $\lambda_s$ is slightly lower than $\lambda_s^m$, there are two segments for the s+p solution. When $\lambda_s$ increases, both the two segments of s+p solution become wider, and the two critical points for p-wave order join together at $\lambda_s=\lambda_s^m$. There should also exist a critical point for the s-wave order for the right segment of s+p solution. But because the critical point for the s-wave order would not be changed by different values of $\lambda_s$, this critical point should stay at $\mu=\infty$, just as the case with $\lambda_s=0$.

We can see that the dashed black line also show a maximum value of $\lambda=\lambda_s^I$. This maximum implies that when $\lambda_s$ is slightly smaller than $\lambda_s^I$, there are two intersection points for the grand potential curves of the s-wave and p-wave solutions. According to the experience that the coexistent solution exit near the intersection point of the grand potential curves for single condensate solutions, the non-monotonic property for the dotted black line is the reason for the non-monotonic property for the solid red line.

We also draw the $\lambda_p-\mu$ phase diagram with $\lambda_s=\lambda_{sp}=0$ in the right plot of Figure.~\ref{fig4}. In this plot, the solid orange line and dashed purple line are the same as in the left cousin. The influence of $\lambda_p$ on the phase structure is similar to the influence of $\lambda_s$, and the increasing of $\lambda_p$ enlarges the region of s+p phase. In contrast, when $\lambda_p$ increases, the left critical point for the s-wave order of the s+p solution moves leftwards while the right critical point for the p-wave order of the s+p solution do not move and the solid red line is vertical. The same argument explains why the right critical point do not move with an increasing $\lambda_p$. The solid blue curve for the left critical point asymptotes to the dashed purple line, which might also be a result of that the dashed black line asymptotes to the dashed purple line.

As the two quartic terms with coefficients $\lambda_s$ and $\lambda_p$ are potential terms in the lagrangian, we focus on the positive values for $\lambda_s$ and $\lambda_p$ in this section. The minus value for the two parameters might cause runaway instability~\cite{Herzog:2010vz}. However, we show the two interesting phase diagrams with negative values of $\lambda_s$ and $\lambda_p$ in Appendix~\ref{Appendix} for possible interest.
\subsection{Tuning $\lambda_{sp}$ and the first order phase transition}
\begin{figure}
\centering
\includegraphics[width=0.6\columnwidth,origin=c,trim=60 260 120 270,clip]{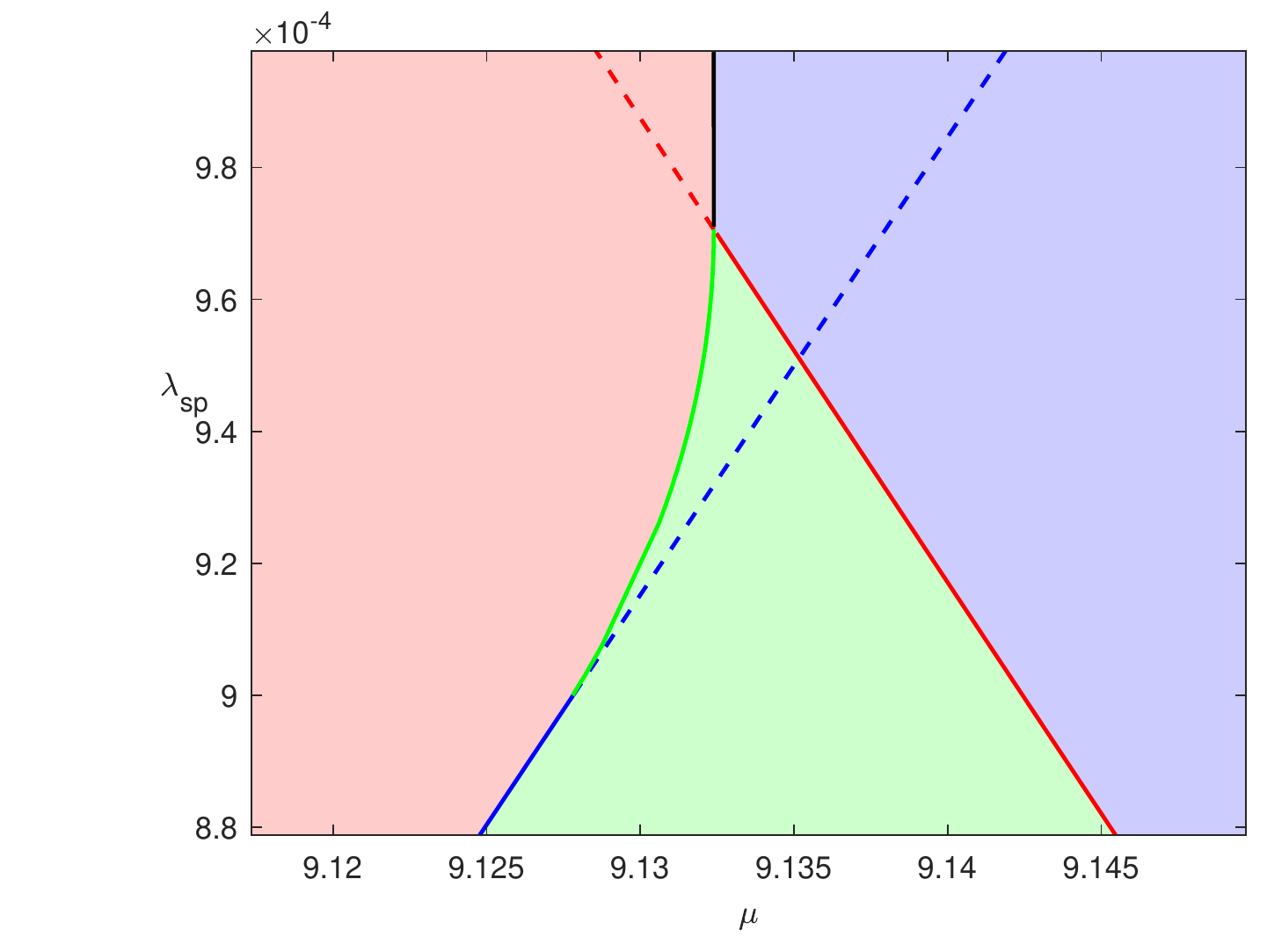}
\caption{The $\lambda_{sp}-\mu$ phase diagram with $\lambda_s=\lambda_p=0$. The blue, red and green region are dominated by the s-wave phase, the p-wave phase and the coexistent s+p phase, respectively. The blue line represents the critical points for the s-wave order of the s+p phase, the red line represents the critical points for the p-wave order of the s+p phase. The solid black line represents the first order phase transition points between the s-wave phase and the p-wave phase, while the solid green line represents the first order phase transition points between the p-wave phase and the s+p phase.}\label{DPLsp}
\end{figure}

The influence of an interaction term such as the one with coefficient $\lambda_{sp}$ in this paper has already been studied in previous work~\cite{Nishida:2014lta,Li:2020ayr}.
Same to the previous results, the term with $\lambda_{sp}$ do not change the single condensate solutions, therefore we focus on the small region near the tri-critical point. We show the $\lambda_{sp}-\mu$ phase diagram with $\lambda_s=\lambda_p=0$ for this model in Figure.~\ref{DPLsp}. In this plot, we use blue, red and green to denote the region dominated by the s-wave phase, the p-wave phase and the s+p phase, respectively. The blue and red lines represent the critical points for the s-wave and p-wave orders of the s+p phase, respectively. The solid section of the two lines are also second order phase transition points, while the dashed lines are not real phase transition points, because the phase transition becomes first order. The solid black line represents the first order phase transition points between the s-wave phase and the p-wave phase, while the solid green line represents the first order phase transition points between the p-wave phase and the s+p phase.

When $\lambda_{sp}$ is increasing, the two critical points moving towards each other along the red and blue lines, making the coexistent phase shrinking and finally disappearing at a tri-critical point with $\lambda_c\approx0.00097$. After $\lambda_{sp}$ reaches the tri-critical point and goes on increasing, the s+p solution is totally unstable. The two critical points for the s+p solution keep moving along the dashed lines and the unstable s+p solution becomes wider. It turns out that the term with parameter $\lambda_{sp}$ is an effectively repelling interaction between the two condensates.

Note that the intersection point of the blue and red lines are not on the black line. As a result, the phase transition from the p-wave phase to the s+p phase becomes first order near the tri-critical point. This is qualitatively the same as the phase diagram in Ref.~\cite{Li:2020ayr}, but is different with that in Ref.~\cite{Nishida:2014lta}, where the phase transition points between the single condensate phases and the coexistent phase are always seconde order. We believe that the detailed small section of first order phase transition are omitted in this earliest result on the topic~\cite{Nishida:2014lta} because of the sparse numerical data points. The tri-critical point without first order phase transition to the coexistent solution means a highly coincident flat plain in the grand potential landscape.
\subsection{reentrant phase}
From the three phase diagrams, we are clear about the influence of three parameters $\lambda_{s}$, $\lambda_{p}$ and $\lambda_{sp}$ and get more power on tuning the phase transitions between the normal phase and three condensate solutions. In this section, we study how to find a reentrant phase transition as a test of the new power.

Since we have fixed $m_s^2=-2, m_p^2=0$ and $q_s=1, q_p=0.8881$, we get the two values of critical chemical potential as $\mu_s=4.06$ for the s-wave solution and $\mu_p=3.88$ for the p-wave solution from the results in Figure.~\ref{Fig1}. Because the three parameters $\lambda_{s}$, $\lambda_{p}$ and $\lambda_{sp}$ do not change the critical points of the single condensate solutions, $\mu_s$ and $\mu_p$ would not change and we always have $\mu_p<\mu_s$. Therefore the system always take a second order phase transition from normal phase to the p-wave phase, and the grand potential curve of the p-wave solution is always lower than the s-wave solution in small $\mu$(left) side.

From the experience of previous studies, the grand potential line for the p-wave solution should be lower both in the left and right side, then it is possible for the system to transform from the p-wave solution to s+p solution at a small value of $\mu$ and reenter the p-wave solution at a larger value of $\mu$ again. In the case with $\lambda_s=\lambda_p=\lambda_{sp}=0$, the grand potential curves of the two single condensate solutions have only one intersection point. Therefore we need to tune some parameter to change the relation in the large $\mu$(right) region. From the influence of $\lambda_s$ and $\lambda_p$ on the grand potential curve shown in Figure~\ref{fig2} and Figure~\ref{fig3}, we can see that we need increase $\lambda_s$ to make the grand potential curves for the s-wave and p-wave solutions have two intersection points.

The dashed black line in the phase diagram in Figure~\ref{fig4} show the trace of the intersection point. We can see from the left plot of Figure~\ref{fig4} that when $\lambda_s$ approaches the maximum value $\lambda_s^I$ for the dashed black curve, we get two intersection points for the two grand potential curves and the grand potential of the p-wave solution is lower than that of the s-wave solution both in the left and right side as we want.

However, the $\lambda_s-\mu$ phase diagram show that in the case of $\lambda_s$ approaches $\lambda_s^I$, the s+p phase dominates the right side of the phase diagram, and we didn't see the reentrance back to the p-wave solution. At this time, let's see the $\lambda_{sp}-\mu$ phase diagram, and remember that increasing the value of $\lambda_{sp}$ reduces the width of the s+p solution. If we want to make the p-wave solution to be more stable than the s+p solution in the right side, we just increase $\lambda_{sp}$. As a result, the two critical points for the s+p solution will move towards each other and finally, we can get the reentrant phase transition. One possible issue is that when we fix the value of $\lambda_s$ and increase $\lambda_{sp}$, the two critical points for the p-wave order of s+p solution may emerge from the center, and then the s+p solution is divided into two segments. This problem can be easily solved by slightly increase the value of $\lambda_s$, which make the two critical points moving towards each other and disappear after the two merge in the middle, as shown by the solid red curve in the left plot of Figure~\ref{fig4}.

\begin{figure}
\centering
\includegraphics[width=0.8\columnwidth,origin=c,trim=90 260 120 280,clip]{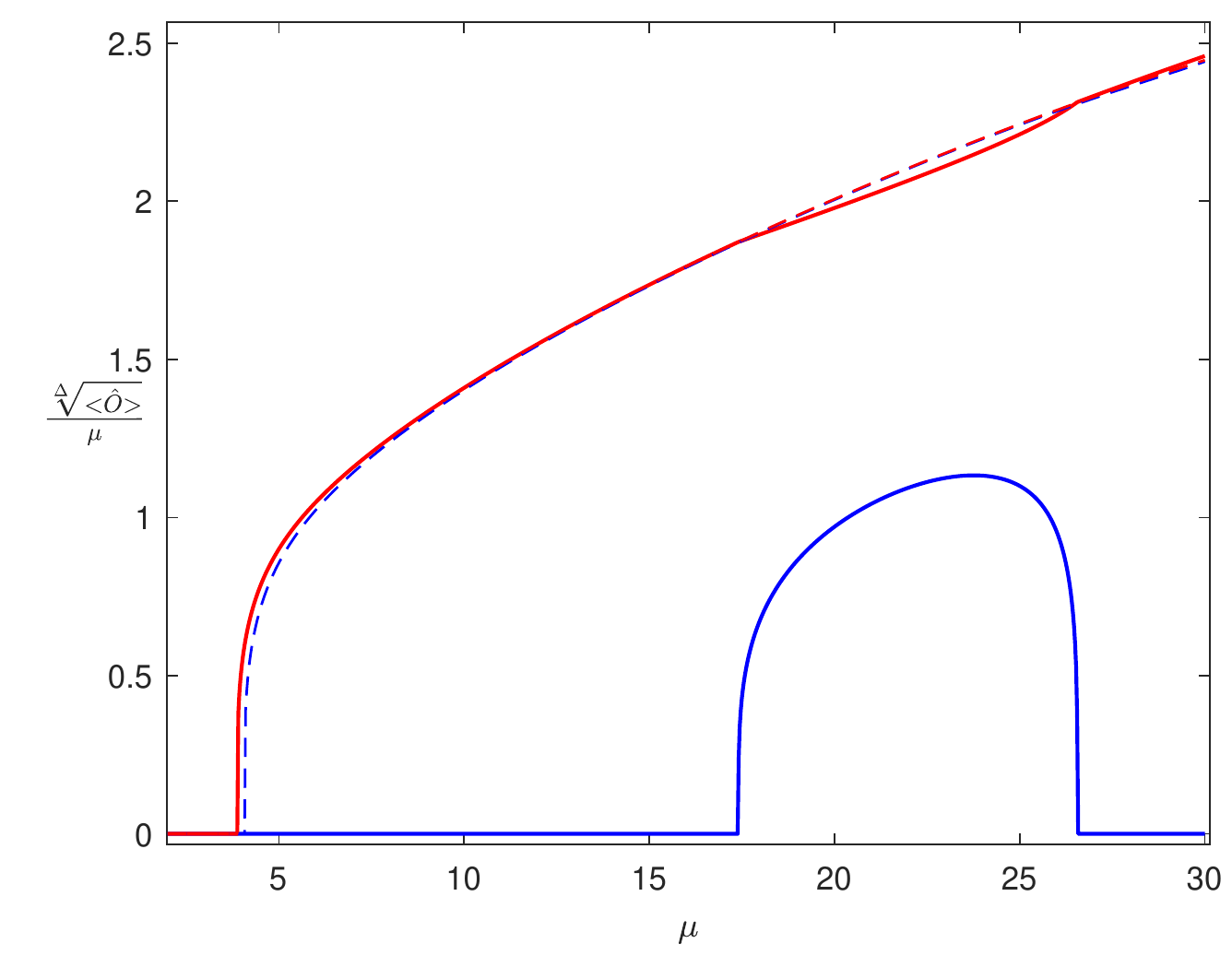}\\
\caption{Condensates of a reentrant phase transition with $m_s^2=-2, m_p^2=0, q_s=1, q_p=0.8881$ and $\lambda_s=0.035, \lambda_p=0, \lambda_{sp}=0.0168$. The red line denotes the condensate of the p-wave order while the blue line denotes the condensate of the s-wave order. The solid lines are for the most stable solutions, while the dashed lines are for the unstable segments of the single condensate solutions.}\label{6}
\end{figure}

Finally, we get a very nice reentrant phase transition with $\lambda_s=0.035, \lambda_{sp}=0.0168$ and $m_s^2=-2, m_p^2=0, q_s=1, q_p=0.8881$, and we show the condensates in Figure.~\ref{6}. We can see in this case that if we increase the value of $\mu$ in the normal phase, the system firstly take a phase transition from the normal phase to the p-wave phase. Then it take the second phase transition from the p-wave phase to the s+p phase. Finally, it takes the third phase transition from the s+p phase back to the p-wave phase. All the three phase transitions are second order, and the reentrance are realized accompanied by the non-monotonic behavior for the s-wave order.

\section{conclusion and discussion}\label{sect:conclusion}
In this paper, we add three quartic nonlinear terms to the lagrangian of the holographic s+p model with two orders. We investigated the influence of the three parameters $\lambda_s$, $\lambda_p$ and $\lambda_{sp}$, respectively, and show the three phase diagrams. With these three terms in hand, we get more power on tuning the phase transitions, and we successfully realized a reentrant phase transition to show this new power.

It should be noticed that, if we take two s-wave order with the same value of $m_s^2$, we are not expected to get the reentrant phase transition. This is because that we can change one s-wave solution with $q_s=q_1$ to the other s-wave solution with $q_s=q_2$ by using the scaling symmetry. Therefore the two solutions with $\lambda_s=0$ are ``parallel'' to each other, and tuning $\lambda_s$ can not make the grand potential curves having two intersection points. This is why we use a s+p model instead of a more simpler s+s model.

Our results on the influence of these parameters are expected to be qualitatively universal in many holographic models. However, some special gravitational theories such as 4D and higher dimensional Einstein-Gauss-Bonnet gravity~\cite{Glavan:2019inb,Fernandes:2020rpa,Qiao:2020hkx} may cause some difference, which is worth being tested in future study. Moreover, our study provides a systematic and convenient way to get some specific phase transitions for future studies on dynamical evolution as well as possible universality. We also expect our work to promote studies on other related topics.

\section*{Acknowledgments}
ZYN would like to thank Li Li for useful discussions. He would also like to thank the organizers of ``2021 Xi'an Conference on Gravitation and Cosmology'' for their hospitality. This work is partially supported by NSFC with Grant No.11965013, 11565017. ZYN is partially supported by Yunnan Ten Thousand Talents Plan Young \& Elite Talents Project.

\appendix
\section{The phase diagrams with negative values of $\lambda_s$ and $\lambda_p$} \label{Appendix}
When the value of $\lambda_s$(or $\lambda_p$) become negative, due to the contribution of the quartic term $-\lambda_s \psi_s^4$(or $-\lambda_p \psi_p^4$), the holographic system will get lower value of grand potential with a larger value of condensate. This implies that the system might be unstable because that the grand potential is not bounded from below. However, we can still get solutions with relatively small condensate value which is at least meta stable and we show the related ``phase diagrams'' in this appendix.

\begin{figure}
\centering
\includegraphics[width=0.47\columnwidth,origin=c,trim=90 260 110 270,clip]{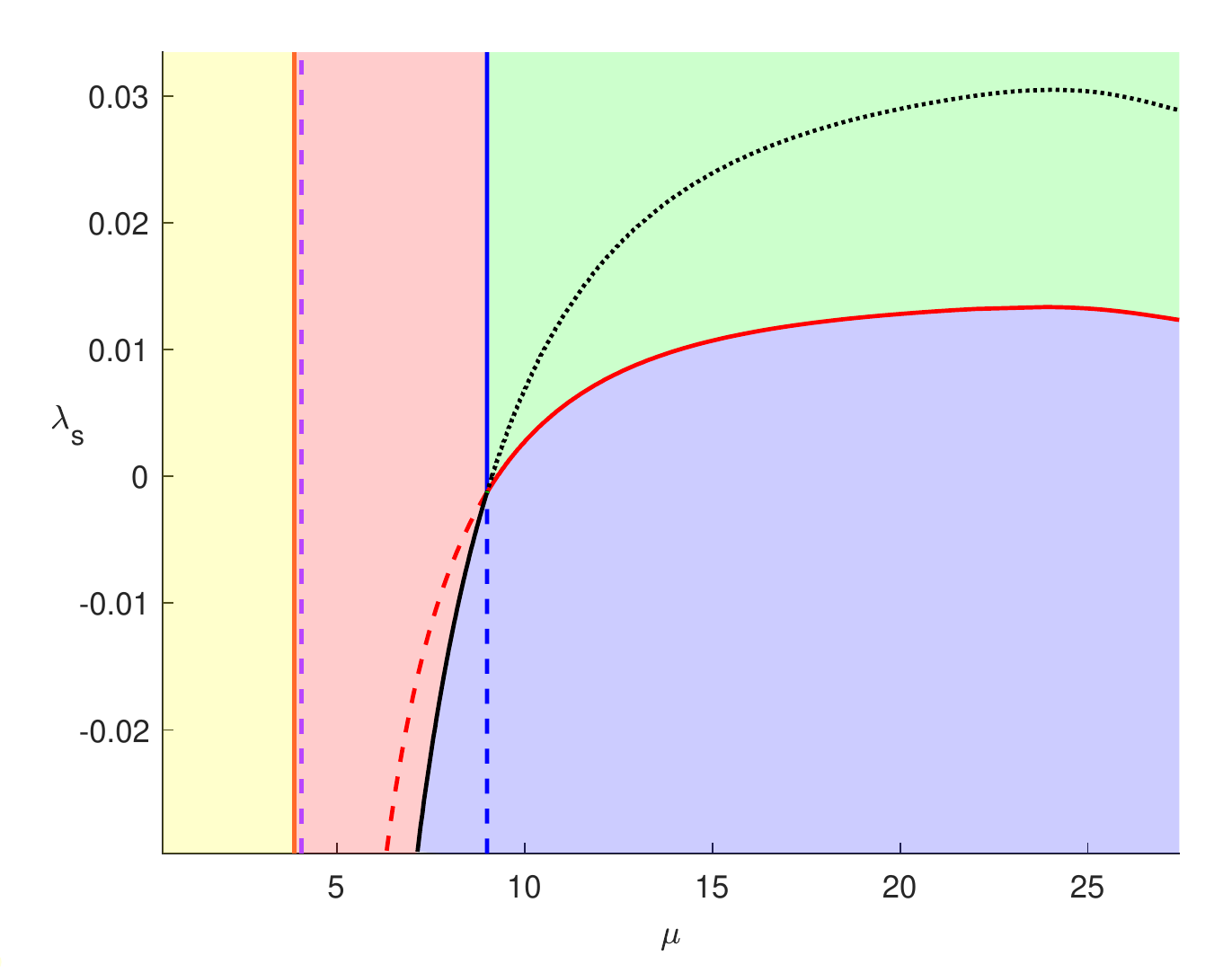}
\includegraphics[width=0.47\columnwidth,origin=c,trim=90 260 110 270,clip]{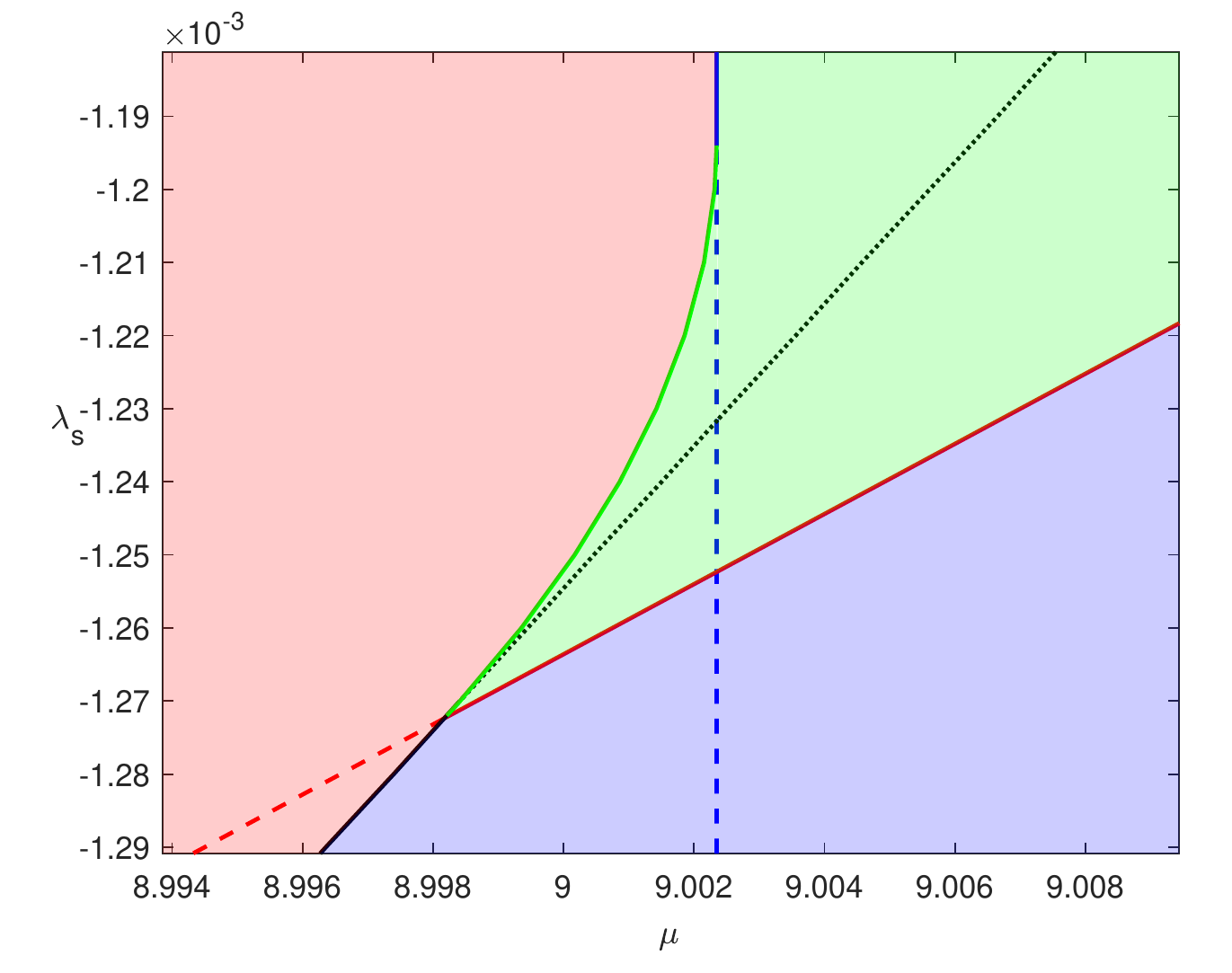}\\
\caption{The $\lambda_s-\mu$ phase diagram with $\lambda_p=\lambda_{sp}=0$ including the negative values of $\lambda_s$. The right plot is an enlarged version near the intersection points of the red, blue and black lines in the left plot. The solid green line denotes the first order phase transition points between the p-wave phase and the s+p phase, while the solid black line denotes the first order phase transition points between the p-wave phase and the s-wave phase. The dashed parts of the blue and red lines indicates that the critical points become spinodal points in first order phase transitions. The other notions are the same as that in the phase diagram with positive values of $\lambda_s$ in Figure~\ref{fig4}.
}\label{negLsmu}
\end{figure}

We extend the $\lambda_s-\mu$ phase diagram with $\lambda_p=\lambda_{sp}=0$ to negative values of $\lambda_s$ in Figure.~\ref{negLsmu}. We can see that the solid blue line is still vertical as we have explained, while the solid red line and the dashed black line go leftwards when the value of $\lambda_s$ decreases. Then the solid blue line, the solid red line and the dashed black line intersect near a small region. Near this small region, we can see that although the critical point denoted by the solid blue line do not move, when the phase transition become first order, the phase boundary between the red and green region move leftwards along the solid green line denoting the phase transition points. When the value of $\lambda_s$ is small enough, the region of s+p phase disappear, and there is a first order phase transition between the s-wave phase and the p-wave phase marked by the solid black curve.

\begin{figure}
\centering
\includegraphics[width=0.47\columnwidth,origin=c,trim=90 260 110 270,clip]{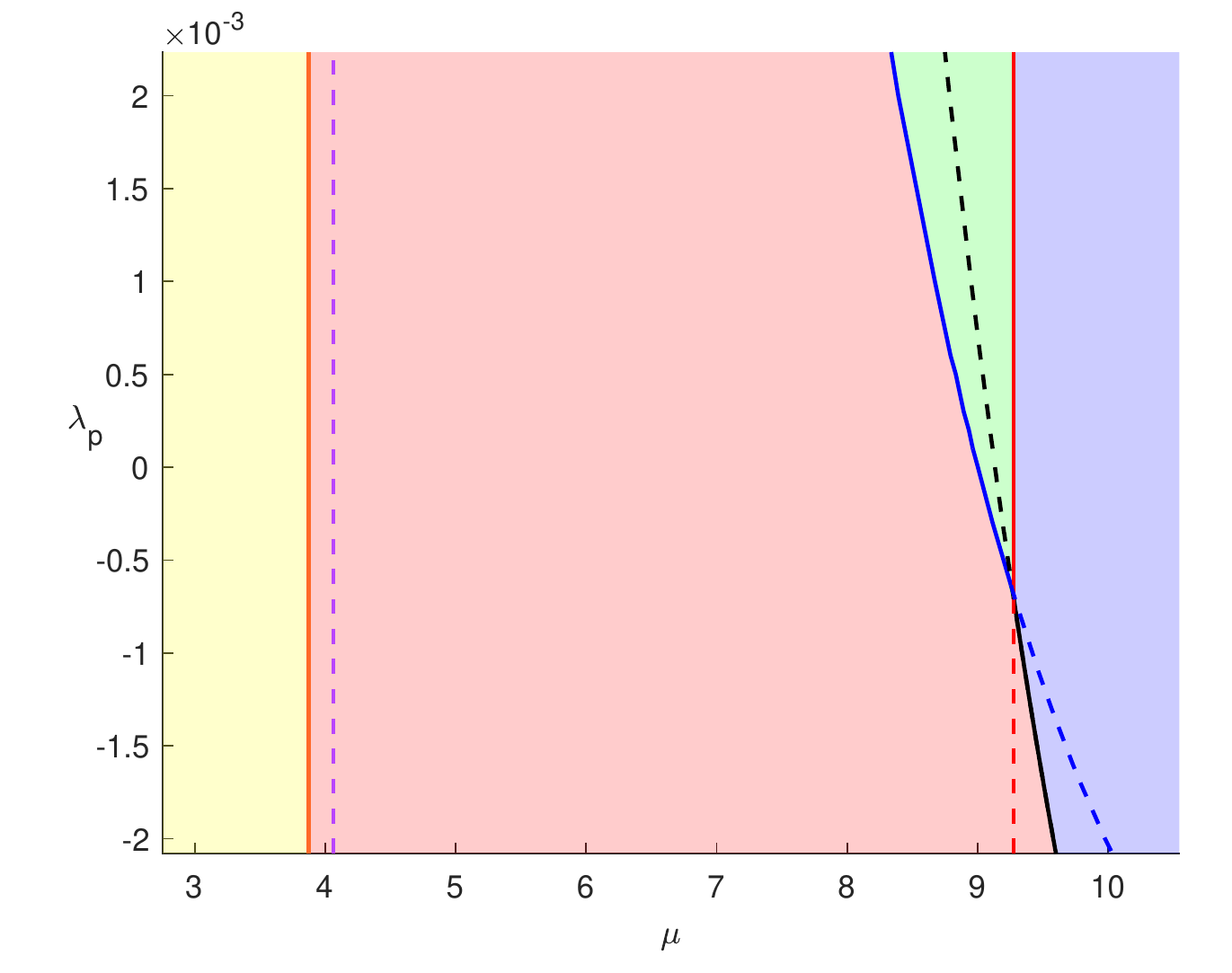}
\includegraphics[width=0.47\columnwidth,origin=c,trim=90 260 110 270,clip]{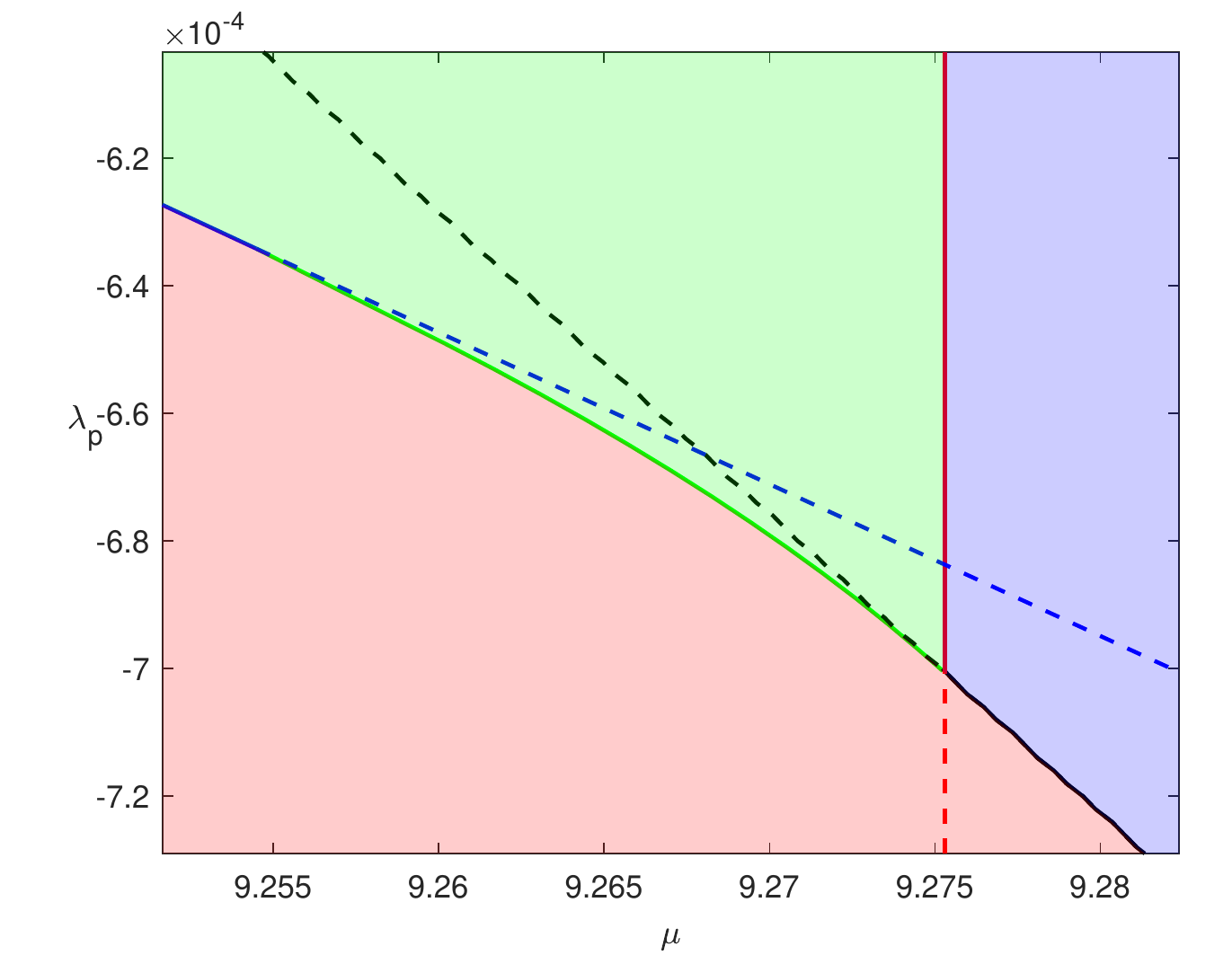}\\
\caption{The $\lambda_p-\mu$ phase diagram with $\lambda_s=\lambda_{sp}=0$ including the negative values of $\lambda_p$. The right plot is an enlarged version near the intersection points of the red, blue and black lines in the left plot. The solid green line denotes the first order phase transition points between the p-wave phase and the s+p phase, while the solid black line denotes the first order phase transition points between the p-wave phase and the s-wave phase. The dashed parts of the blue and red lines indicates that the critical points become spinodal points in first order phase transitions. The other notions are the same as that in the phase diagram with positive values of $\lambda_s$ in Figure~\ref{fig4}.
}\label{negLpmu}
\end{figure}

The case for the $\lambda_p-\mu$ phase diagram is similar. We also extend the $\lambda_p-\mu$ phase diagram with $\lambda_s=\lambda_{sp}=0$ to negative values of $\lambda_p$ in Figure.~\ref{negLpmu}. We can see that the solid red line is vertical, while the solid blue line and dashed black line move rightwards when $\lambda_p$ decreases. The three lines also intersect in a small region. Near this small region, the phase transition from the p-wave phase to the s+p phase becomes first order and we use solid green line to denotes the phase transition points. Although the red line denoting the critical points for the p-wave order of the s+p phase is always vertical, the s+p phase disappear when $\lambda_p$ is low enough and there is a first order phase transition between the p-wave phase and the s-wave phase marked by the solid black curve.

We should notice that the system might suffer from runaway pathologies~\cite{Herzog:2010vz} dual to the negative contribution of the quatic terms when $\lambda_s$ or $\lambda_p$ is negative. However, we still extend the phase diagram for the condensed solutions to negative values of $\lambda_s$ or $\lambda_p$. We will further investigate the related stability problem in future study. In this paper, we focus on the phase diagram and reentrant phase transitions with positive values of $\lambda_s$ and $\lambda_p$ in the main text, which can be applied in further studies with out worrying about the stability problem.

\end{document}